\documentclass[epj]{svjour}
\usepackage{graphicx}
\usepackage{mathtools}
\usepackage{subfigure}
\usepackage{graphics}
\usepackage{amssymb,amsmath}
\usepackage{color}

\usepackage{ulem}

\begin{document}
\title{Universality in Ground State Masses of Nuclei}

\author{A. Bhagwat\inst{1} and Sudhir Ranjan Jain\inst{2,3,4}}

\institute{School of Physical Sciences, UM-DAE Centre for Excellence in Basic Sciences, University of Mumbai, 
Vidyanagari Campus, Mumbai 400 098, India \and
Theoretical Nuclear Physics and Quantum Computing Section, Nuclear Physics Division, Bhabha Atomic Research Centre, Mumbai 400085,
India \and
Homi Bhabha National Institute, Training School Complex, Anushakti Nagar, Mumbai 400 094, India \and
UM-DAE Centre for Excellence in Basic Sciences, University of Mumbai, Vidyanagari Campus, Mumbai 400 098, India}
%
%

\date{Received: date / Revised version: date}
%

\abstract{
The beautiful and profound result that the first eigenvalue of Schr\"odinger operator 
can be interpreted as a large deviation of certain kind of Brownian motion leads to 
possible existence of universality in the distribution of ground state energies of 
quantal systems. Existence of such universality is explored in the distribution of 
the ground state energies of nuclei with $Z\ge8$ and $N\ge8$. Specifically, it has 
been demonstrated that the nuclear masses follow extreme-value statistics, implying 
that the nuclear ground state energies indeed can be treated as extreme values in the 
sense of the large deviation theory of Donsker and Varadhan.
}
\maketitle

\maketitle


\section{Introduction}
A glimpse at the energy levels of nuclei reveals their complexity: the ``bar codes'' 
seem to bear similarity at a gross level with some nuclei showing more clustering than 
others. Larger clustering is related to degeneracies and well-known magic numbers in 
nuclei \cite{bm} and metallic clusters \cite{brack}, with origin in underlying symmetries. 
Energy level patterns are characterized at a gross level in terms of mean density, 
epitomized in Thomas-Fermi and other semiclassical formulae. A closer look brings out 
fluctuation properties which correlate densities at two or more energies \cite{ap,guhr,kota}. 
In the scaling limit, the level correlations have been successfully explained by Random Matrix Theory (RMT). 
However, at the centre of the physics of nuclei resides the ground states and the corresponding 
nuclear masses. Here we address ``the elephant at the centre'' and present a compelling case for 
universality associated with ground state energies. Its origin is in our consideration of ground 
state (GS) energies as extreme values \cite{comment} of the spectra. The collection of these 
extreme values, taken for all nuclei, constitutes an ensemble, named as ground state ensemble (gSE). 

The GS corresponds to the first eigenvalue of the self-adjoint operator, which is calculated 
by setting up a variational formulation leading to the Rayleigh - Ritz formula. Ground state 
energy for a Hamiltonian operator may also be calculated by employing the Feynman - Kac formula which 
is based on the asymptotic form of the Green function. It turns out that the first eigenvalue of 
the Schr\"{o}dinger operator can be interpreted as a large deviation of certain kind of a Brownian 
motion. Donsker and Varadhan generalized this variational formulation to what they call the 
principal eigenvalue (GS in our nomenclature) for operators with maximum principle \cite{dv}, 
thus providing a large-deviation interpretation of variational formula which reduces to the 
Rayleigh - Ritz formula for self-adjoint operators. This beautiful work is at the basis of our result.  

\section{Computational Details and Results}
The goal of this work is to establish the universality argued above, which is 
associated with the ground state energies of nuclei quantitatively. For this purpose, we choose 
nuclei with atomic number, $Z\ge8$ and neutron number, $N\ge8$. For all 2353 nuclei, experimental 
masses have been considered from the atomic mass evaluation of 2012 \cite{WAN.12a,WAN.12b}. The 
fluctuations in the ground state energies can be obtained by subtracting the liquid drop binding 
energies ($E_{\rm LDM}$) from the experimental binding energies. This procedure is justified, and 
can be argued to be robust, since it is well known that all the variants of the liquid drop model 
yield binding energies that are similar to each other. Thus, the general conclusions drawn are not 
sensitive to a particular choice of the liquid drop model. Here, we consider a liquid
drop model \cite{POM.03,BHA.12a,BHA.12b,BHA.21} with binding energy $E_{LDM}$ (see Appendix A for details). 

\begin{figure}[htb]
\centering \includegraphics[scale=0.32]{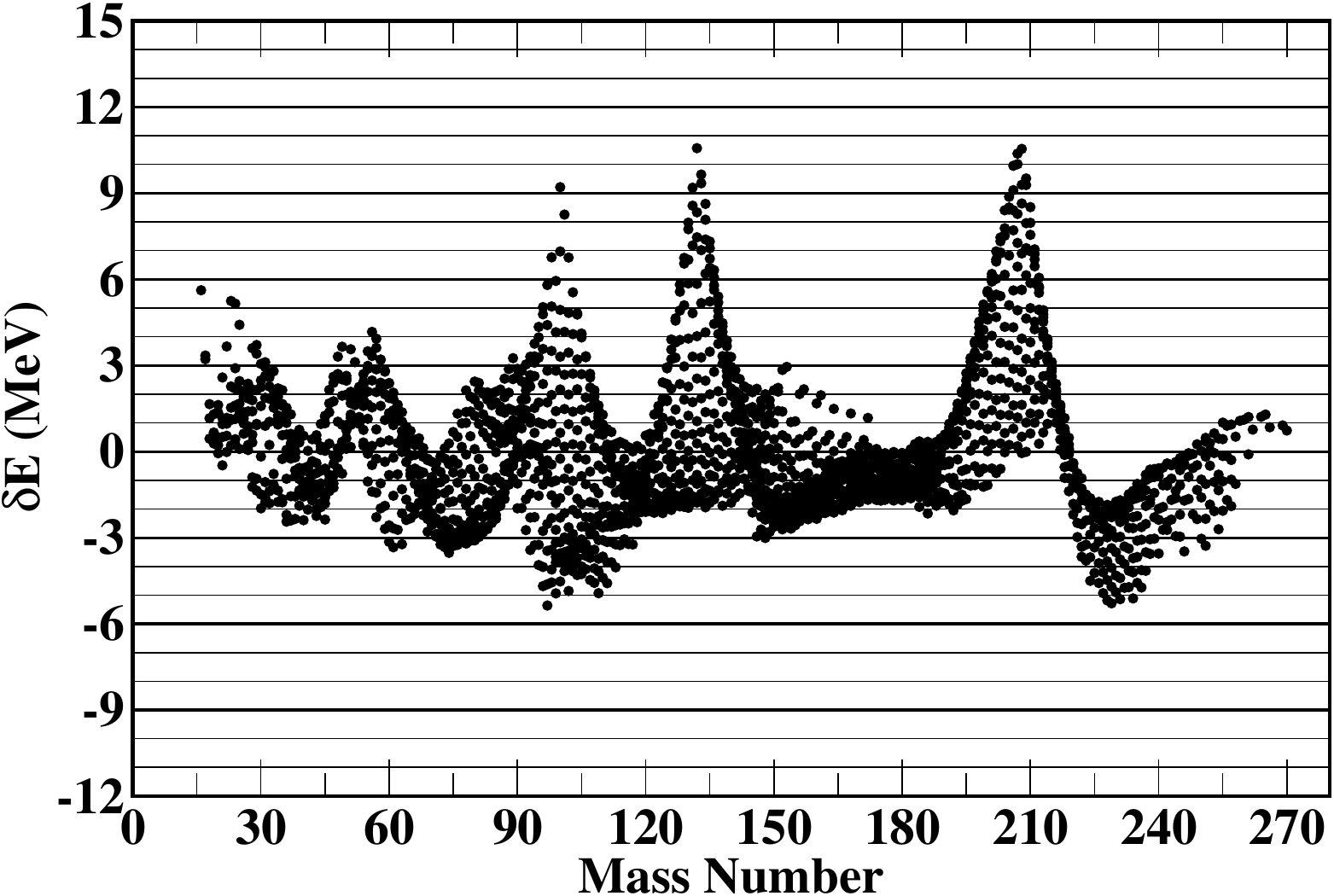}
\caption{Difference between experimental and LDM binding energies. These are sharply peaked at the
magic nuclei such as $^{100,132}Sn$ and $^{208}$Pb. }
\label{diffbe}
\end{figure}

We denote the difference between experimental energies and $E_{LDM}$ by $\delta E$, and
it is this quantity that we shall be analysing further to investigate the  universality
associated with the ground state energies. Here, we adopt the sign convention where
the quantity $\delta E$ is positive at shell closures. Notice that the LDM assumed above 
does not have any deformation dependence, nor it contains any terms such as the Wigner 
correction. The only effect that has been taken in addition to the usual `macroscopic' 
terms is the pairing energy. This is justified again on the basis of an idea similar to 
the Strutinsky theorem: exactly like the binding energy (barring two - body correlations), 
the pairing energy can be thought to be made of a `smooth' part and a fluctuating part \cite{VIN.11}.

The quantity $\delta E$ is plotted in Fig. (\ref{diffbe}) for the chosen set of 2353 nuclei 
as a function of mass number. It can be seen that the sharp peaks are observed precisely at 
the locations of doubly magic nuclei. The mean of $\delta E$ turns out to be nearly zero. 
The plot also hints towards the possible choice of bin size for statistical study of this 
data set: an inspection of the graph suggests a bin size of about 1 MeV. For reliable conclusions, 
the bin size has to be optimal, the analysis of which we come to now. The question of optimality 
can be answered by a binning method developed by Shimazaki and Shinomoto \cite{SHI.07}. The 
idea is to choose the bin size such that the cost function is minimized, the details of this 
well-known procedure are explained in Appendix B. 

Using the present data, Shimazaki-Shinomoto procedure gives an optimal bin-width to be $h \sim 1$ MeV. 
\begin{figure*}[htb]
\centering \hbox {\includegraphics[scale=0.34]{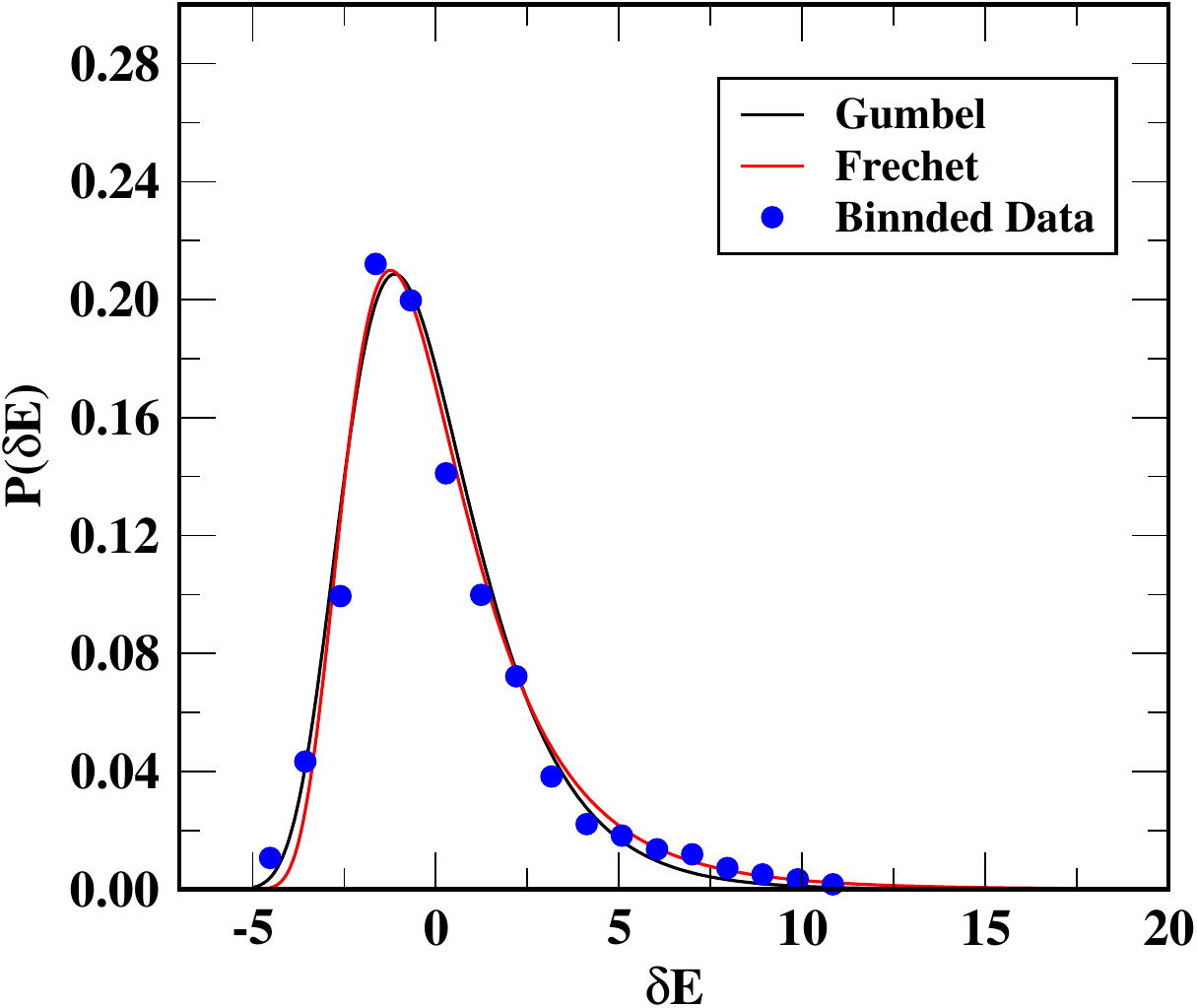} \hfill \includegraphics[scale=0.34]{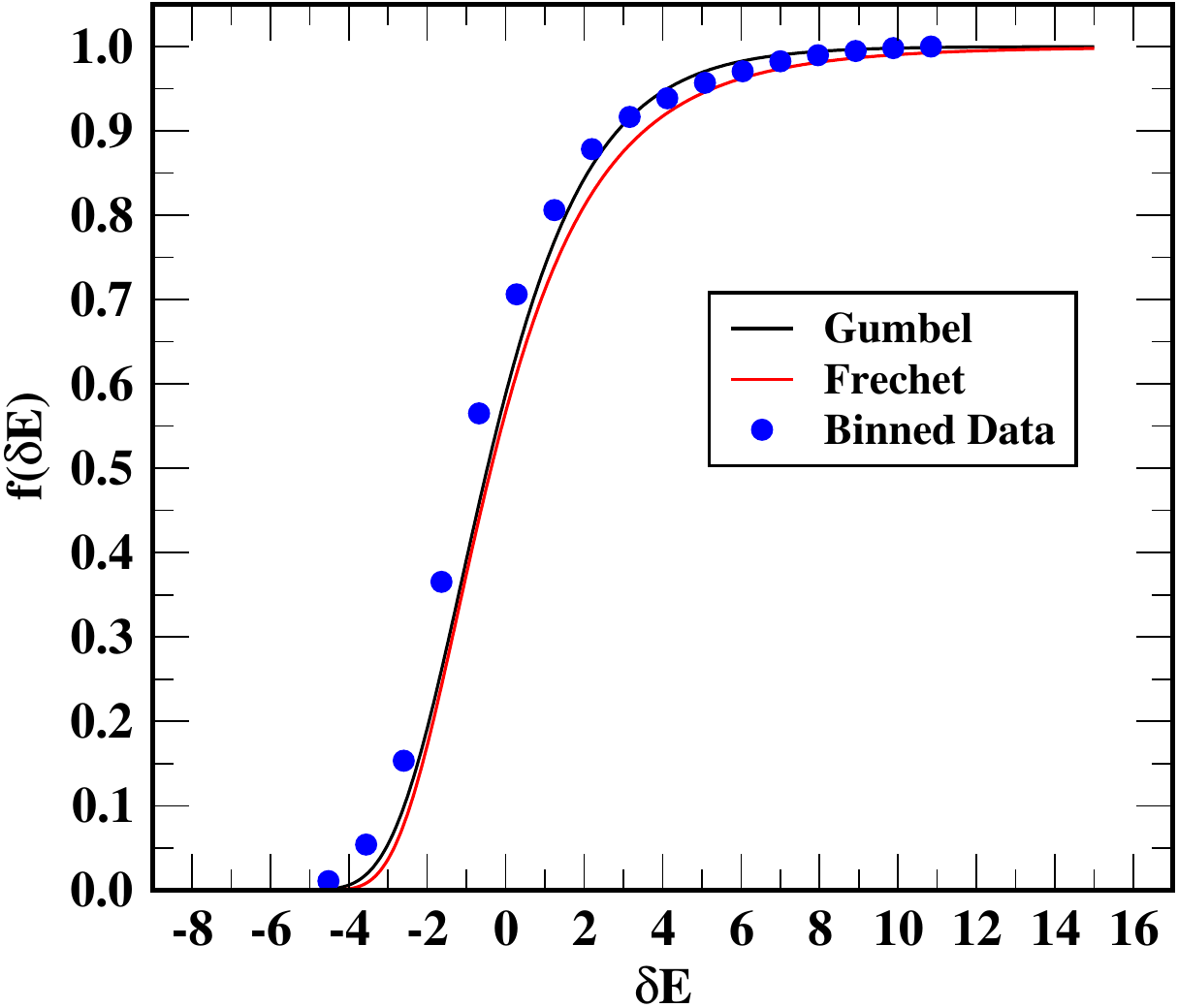}}
\centering \hbox {\includegraphics[scale=0.34]{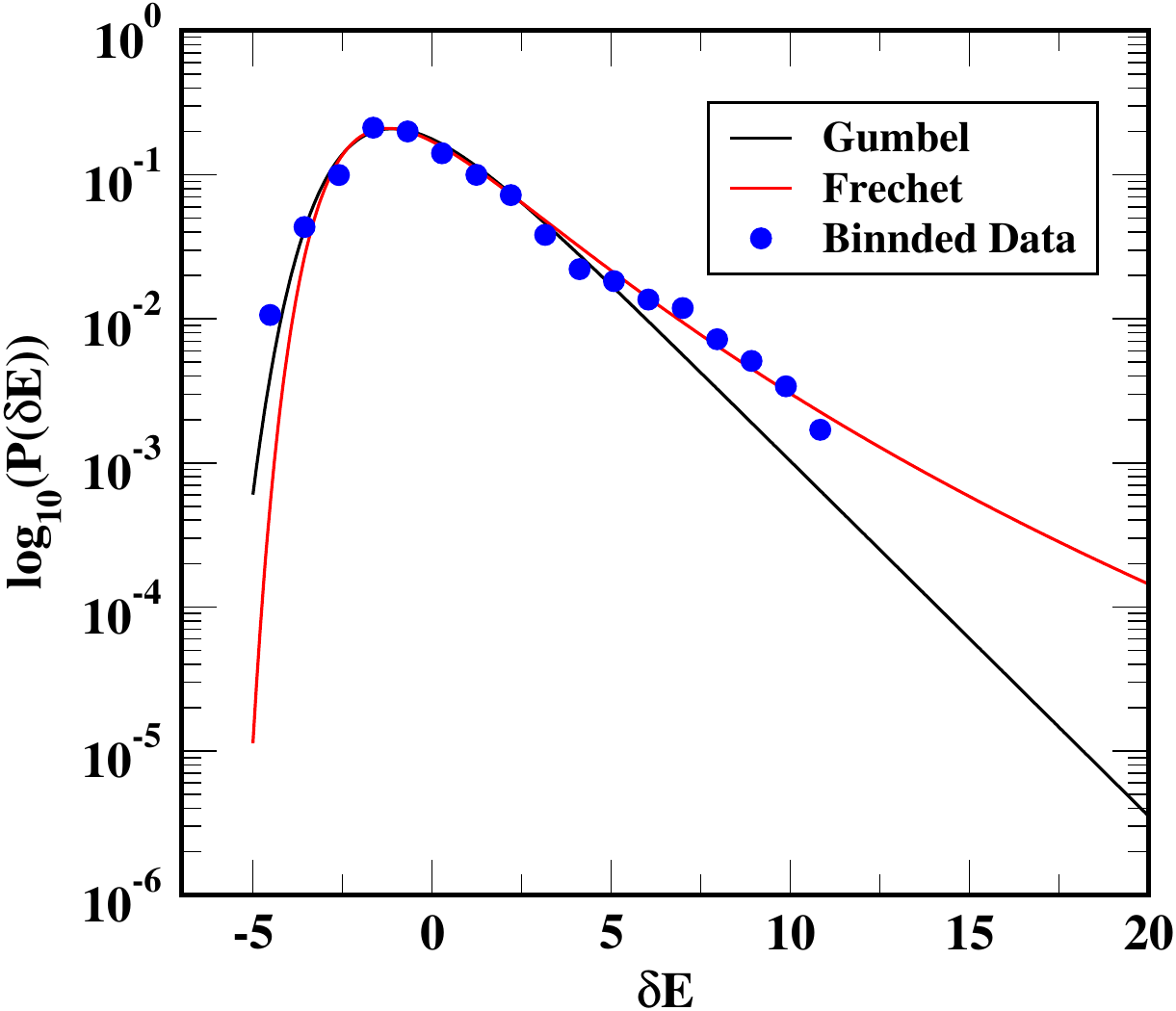}\hfill
\includegraphics[scale=0.34]{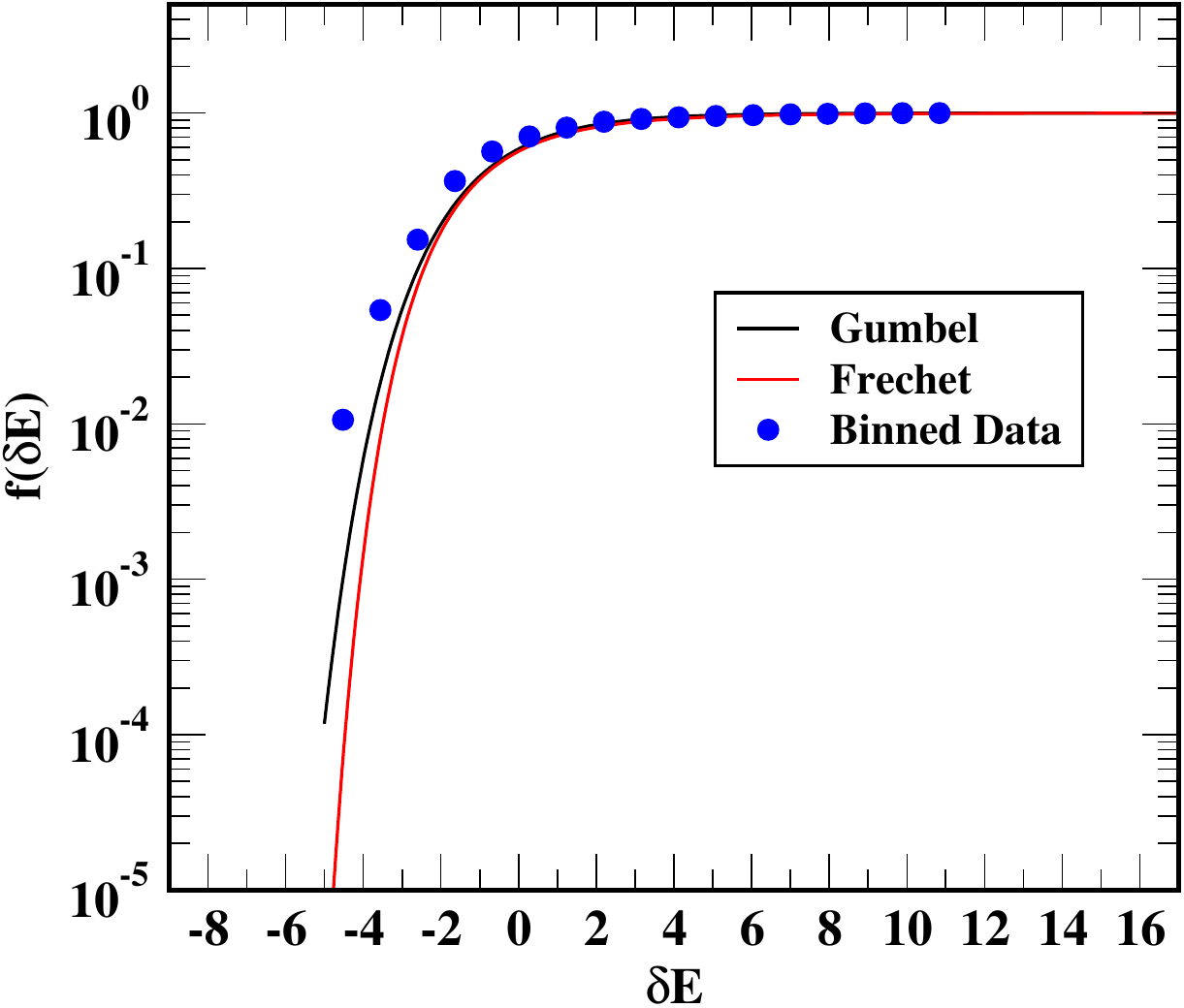}}
\caption{The probability distribution $P(\delta E)$ obtained from binning procedure as a function 
of $\delta E$. Fits to the Gumbel and Fr\'echet distributions and cumulative distributions on linear 
as well as $\log$ scale are shown.}
\label{dist}
\end{figure*}
The resulting frequency distribution, normalized to unity, is plotted in Fig. \ref{dist}. The
distribution is clearly different from Gaussian. In the light of the argument suggesting the possibility
of the distribution being one of the extreme-value distributions (GEV), we first check if the 
distribution is a Gumbel distribution, expressed as:
\begin{eqnarray}
P\left(x\right) = \frac{1}{\beta}\exp\left\{ - \frac{\left(x - \mu\right)}{\beta} - \exp\left[- \frac{\left(x - \mu\right)}{\beta} \right] \right\}
\end{eqnarray}
where the scale parameter $\beta\in\mathbb{R}^{+}$ and the location parameter $\mu \in \mathbb{R}$.
The support of this distribution is the whole of $\mathbb{R}$, i.e., $x \in \mathbb{R}$. The
cumulative distribution function is expressed as:
\begin{eqnarray}
f\left(x\right) = \exp\left\{ - \exp\left[- \frac{\left(x - \mu\right)}{\beta} \right] \right\}.
\end{eqnarray}
The parameters
$\mu$ and $\beta$ in this work are obtained through $\chi^2$ minimisation, using the
well-known Powell's conjugate direction method \cite{POW.64}, which is a derivative - free optimization
technique for finding local minima of a given function. The values of the parameters thus 
obtained are $\mu$ = -1.12445 and $\beta$ = 1.78709. The fit turns out to be good, with
an {\it r.m.s.} deviation of 1.4$\times10^{-4}$. 

The observation recorded above needs to be substantiated further. In order to do that, 
we consider the general GEV distribution:
\begin{eqnarray}
P\left(x\right)&=&\frac{1}{\beta}\left\{1 + \xi\left(\frac{x - \mu}{\beta}\right)\right\}^{-\left(1 + \xi\right)/\xi} \times
\nonumber \\
&& \hspace{48pt}\exp{\left\{1 + \xi\left(\frac{x - \mu}{\beta}\right)\right\}^{-1/\xi}}
\end{eqnarray}
If $\xi$ = 0, we get the Gumbel distribution and the corresponding support is $\mathbb{R}$. 
On the other hand, if $\xi < 0$, one obtains the Weibull distribution, with the support that 
is bounded above, that is, with $x \in \left(-\infty\right., \left.\mu - \beta/\xi\right]$.
Finally, if $\xi > 0$, the Fr\'echet distribution is obtained, with the support that is bounded below,
such that $x \in \left[\mu - \beta/\xi\right., \left.\infty\right)$. The cumulative distribution function
in the case of GEV is expressed as:
\begin{eqnarray}
f\left(x\right) = \exp\left\{1 + \xi\left(\frac{x - \mu}{\beta}\right)\right\}^{-1/\xi}
\end{eqnarray}

The parameters are determined using Powell's method. The explicit values of the parameters 
are: $\mu$ = -1.03173, $\beta$ = 1.79009 and $\xi$ = 0.13552 with {\it rms} deviation 
of 1.2$\times10^{-4}$. The fact that $\xi > 0$ indicates that this is a Fr\'echet distribution.

\section{Statistical Analysis}
On the face of it, it seems that we have fitted the same data equally well with two different
probability densities. However, we shall now establish that the two seemingly different distributions
are statistically indistinguishable. To begin with, notice that the values of the
position and scale parameters in the two cases are very close to each other: the position parameters
differ by about 5\%, whereas the scale parameters differ by less than 0.2\%. Secondly, the value of the 
shape parameter is just about 0.14 which is quite small indeed. The lower bound on the support
works out to be $\sim$-14, which is very large and negative (notice that the smallest
value of $\delta E$ in this data set is $\sim$-4.5). Thus, it is to be expected
{\it a priori} that the two should be almost identical, particularly so far as shapes are concerned.
This observation is amply reflected in
Fig. (3), where, we plot the two probability distributions and cumulative distributions along
with the corresponding binned data. It is clear from the graph that the two distributions are
very close to each other, and also describe the binned data very well. The cumulative distribution
however shows that there are differences between Gumbel and Fr\'echet, and we shall quantify
them here.

To quantify this near-indistinguishability of the two distributions, one sample Kolmogorov-Smirnov 
test (see, for example, \cite{DOD.08}) is carried out to test the null hypothesis that the fitted 
Gumbel and Fr\'echet empirical data (cumulative probability at the midpoint of a bin) is identical 
to the one for Gumbel distribution (Eqs. (1) and (2)). The former is denoted by $S_{n}(x)$ and the 
latter by $f(x)$. Then the Chebyshev norm,
\begin{eqnarray}
D = \max_{x\in \mathbb{R}} \left| S_{n}(x) - f(x)\right|, \quad n\in\mathbb{N}
\end{eqnarray}
gives the Kolmogorov - Smirnov statistic. 
Let the significance level of an observed value of $D$ be denoted by $P$. small values of $P$ 
show that the cumulative distribution function $S_n$ is significantly different from the 
hypothesised function, $f(x)$ (for details, refer to \cite{DOD.08,NR.92}). In other words, 
the null hypothesis that the two are the same is to be accepted if the $P$ value is close to 1. 
In the present case, $D$ turns out to be equal to 1.5$\times10^{-2}$, leading to $P\sim 1$, 
confirming the null hypothesis. We now perform the Kolmogorov - Smirnov test once again for the 
cumulative distributions. Even in this case, the quantity $D$ works out to be 3.3$\times10^{-2}$, 
leading to $P\sim 0.98$, supporting the above conclusion. From these two conclusions alone, 
one may conclude that within the Kolmogorov - Smirnov criteria, the two distributions are 
almost identical. 

We now test the closeness of the `observed' distribution (binned data) to the fitted
Gumbel distribution. The null hypothesis, in this case, is that the empirical data (probability at
the midpoint of a bin) is identical to the one for that of the Gumbel distribution. Again,
the infinity norm ($D$) turns out to be $\sim$0.1, leading to $P\sim 0.99$, validating
the null hypothesis. As we had done earlier, we now apply the Kolmogorov - Smirnov
test to the cumulative distributions. In that case, $D$ works out to be $\sim$0.11, leading
to $P\sim0.99$, substantiating the claim that indeed, the distribution is a Gumbel
distribution as per the Kolmogorov - Smirnov test.

We next compute the quantities,
mean ($E[X]$), variance ($\mathrm{Var}(x)$),
skewness ($\tilde{\mu}_3$), excess kurtosis (${\cal{K}}[X]$) and
Shannon entropy ($H(X)$),
characterising the Gumbel and Fr\'echet distribution, as well
as the binned data. The explicit expressions 
for these quantities are summarised in Appendix C.

\begin{table*}[htb]
\caption{Various quantities characterising the 
probability distributions. Here the quantities listed under `Binned Data' stand for those computed
directly from the binned data, `Exact' means the ones calculated from the exact expressions for 
all the relevant quantities (detailed in Appendix C), and `Bin Centroid' implies the quantities computed by taking $\delta E$ to be
the bin centroid.}
\begin{center}
\begin{tabular}{cccccc} \hline
                &               &\multicolumn{4}{c}{From Distribution} \\ \cline{3-6}
                &               &\multicolumn{2}{c}{Gumbel}& \multicolumn{2}{c}{Fr\'echet} \\\hline
Quantity        &   Binned Data &    Exact    & Bin Centroid & Exact   &Bin Centroid   \\\hline
  $E[X]$        &    0.003      &   -0.099    &  -0.111      & 0.222   &  0.152         \\
Var$(X)$        &    6.319      &    5.112    &   5.218      & 7.475   &  6.454         \\
$\tilde{\mu}_3$ &    1.269      &    1.140    &   1.038      & 2.140   &  1.310         \\
${\cal{K}}[X]$  &    2.144      &    2.4      &   1.568      & 10.549  &  2.195          \\
$H(X)$          &    2.247      &    2.144    &   2.228      & 2.215   &  2.270         \\\hline
\end{tabular}
\end{center}
\end{table*}

A comparison between
these quantities (denoted by `From Distribution') and
the corresponding quantities computed directly from binned data (denoted by `Binned Data')
is made in Table (1). In addition, the quantities computed by using the exact
distribution functions, but only the bin centroids are also presented there.
By and large, these agree with each other reasonably well. The only exception is
excess kurtosis, which has been grossly underestimated for the Fr\'echet distribution.
The binned data, exact result for Gumbel, and kurtosis obtained for Fr\'echet from
centroids are quite close to each other, whereas the exact result for Fr\'echet is
way too high: this is to be expected, as can be seen from the plot where the distributions
have been presented on log scale: the Fr\'echet distribution tends to have a rather heavy tail,
and this dominates kurtosis: it is well known that it is both the peak and tail that
contribute to kurtosis (see, for example, \cite{DEC.97}). Fr\'echet has a rather heavy tail, and while computing the
kurtosis from bin centroids (with exact distribution), the tail has not been fully
accounted for.

\section{Discussion and Summary}

The distribution of $\delta E$ is a GEV. As per the Kolmogorov - Smirnov test, as
well as the visual confirmation that we get after examining the two distributions
along with the binned data, it is apparent that the Gumbel and Fr\'echet
distributions here behave alike, given that the parameter $\xi$ is small: it is
a well known fact that in the limit of $\xi \rightarrow 0$, Fr\'echet distribution
`tends' to Gumbel. Thus, so far as these measures are concerned, the two are
almost identical, and the distribution can be safely taken to be a Gumbel distribution.
This is further supported by the observation that the excess kurtosis value of
the binned data is very similar to that of the Gumbel distribution. Even the
excess kurtosis obtained by assuming exact Gumbel distribution and centroids of the
bins yields an excess kurtosis that is not too far from these two. Fr\'echet distribution,
on the other hand, has a very large excess kurtosis, which does not really agree with
the other calculations. The excess kurtosis obtained from bin centroids is closer to
the others, but this is primarily due to the fact that the entire distribution is
not taken into account here.

Whereas the energy levels of nuclei are known to possess correlations which are 
well-understood in terms of RMT \cite{ap,bhp}, we have argued that the ground state 
energies may be treated as extreme values in the sense of large-deviations theory. 
This is not only interesting but also significant as we are able to propose a universality 
for nuclear masses which happens to belong to one of the well-known distributions. We 
would like to emphasize that the kernel of the argument leading to our finding is in 
the profound work by Donsker and Varadhan \cite{dv}. 

\section*{Acknowledgment}
This article is dedicated to the memory of one of the pioneering random matrix theorists 
and a dear colleague, Akhilesh Pandey.  

\appendix
\section{Liquid Drop Model}
Here, we consider a liquid
drop model inspired from Pomorski's work \cite{POM.03,BHA.12a,BHA.12b,BHA.21,BHA.14}, expressed as
\begin{eqnarray}
E_{LDM}&=&a_v\left[1~+~\frac{4k_v}{A^2}~T_z\left(T_z~+~1\right)\right]A \nonumber \\
       &+& a_s\left[1~+~\frac{4k_s}{A^2}~T_z\left(T_z~+~1\right)\right]A^{2/3} \nonumber \\
&+& \frac{3Z^2e^2}{5r_0A^{1/3}}~+~\frac{C_4Z^2}{A} ~+~E_p.
\label{LDM}
\end{eqnarray}
In this expression, $T_z$ is the third component of isospin and $e$ is the electronic charge. The
coefficients $a_v$, $k_v$, $a_s$, $k_s$, $r_0$ and $C_4$ (correction to Coulomb energy due to 
surface diffuseness) are treated to be free parameters. Following M\"oller and Nix \cite{NIX.92}, the smooth 
pairing energy is assumed to be of the form
\begin{eqnarray}
E_p &=& \frac{d_n}{N^{1/3}} ~,~\mathrm{for}~N~\mathrm{odd~and~}Z\mathrm{~even} \nonumber \\
    &=& \frac{d_p}{Z^{1/3}} ~,~\mathrm{for}~Z~\mathrm{odd~and~}N\mathrm{~even} \nonumber \\
    &=& \frac{d_n}{N^{1/3}} + \frac{d_p}{Z^{1/3}} + \frac{d_{np}}{A^{2/3}} ~,~
\mathrm{for}~Z~\mathrm{and~}N\mathrm{~odd} \nonumber \\
    &=& 0~,~ \mathrm{for}~Z~\mathrm{and~}N\mathrm{~even}.
\end{eqnarray}
with the constants $d_n$, $d_p$ and $d_{np}$ being free parameters. These parameters have been determined
through $\chi^2$ - minimisation through the NETLIB \cite{NETLIB} implementation of the 
Levenberg - Marquardt (LM) algorithm \cite{MAR.68,KEL.99}.
The values of the coefficients thus obtained are \cite{BHA.14}:
$a_v$ = -15.505 MeV; $a_s$ = 17.830 MeV; $k_v$ = -1.825;
$k_s$ = -2.265; $r_0$ = 1.215 fm; $C_4$ = 1.297 MeV; $d_n$ = 4.687 MeV; $d_p$ = 4.717
MeV; and $d_{np}$ = -6.495 MeV. The {\it rms} deviation obtained for the fit, as
expected \cite{MYE.96}, is 2.456 MeV.

\section{Optimisation of Bin-size}
Suppose, that for given
width $\Delta$, there are $M$ bins in all. Let $n_k$ represent number of
objects in the $k^{th}$ bin. Then, the mean $\bar{n}$ and variance $v$ corresponding
to the data are defined by obvious expressions \cite{SHI.07} 
\begin{eqnarray}
\bar{n} =\frac{1}{M}\sum_{j=1}^{M}n_k, \qquad
v = \frac{1}{M}\sum_{j=1}^{M}\left(n_k - \bar{n}\right)^2.
\end{eqnarray}
Given these quantities, the cost function, which depends explicitly on the bin width,
is given by \cite{SHI.07}
\begin{eqnarray}
C\left(\Delta\right) = \frac{2 \bar{n} - v}{\Delta^2}.
\end{eqnarray}
The Shimazaki-Shinomoto procedure amounts to determination of $\Delta$ such
that the above cost function is minimized. 

\section{Gumbel and Fr\'echet distribution}

Quantities calculated for Gumbel and Fr\'echet distribution, as well
as for the binned data are detailed here: mean ($E[X]$), variance ($\mathrm{Var}(x)$),
skewness ($\tilde{\mu}_3$), excess kurtosis (${\cal{K}}[X]$) and
Shannon entropy ($H(X)$). These quantities can be computed exactly for a Gumbel as well as a
Fr\'echet distribution (see, for example, 
\cite{FIN.03} for further details).
We first list these quantities for a Gumbel distribution:
\begin{eqnarray}
E[X] &=& \mu + \beta \gamma \\
\mathrm{Var}(X) &=& \frac{\pi^2}{6}\beta^2 \\
\tilde{\mu}_3 &=& \frac{12 \sqrt{6}\,\zeta\left(3\right)}{\pi^3} \\
{\cal{K}}[X] &=& \frac{12}{5} \\
H(X) &=& \log\left(\beta\right) + \gamma + 1
\end{eqnarray}
where, $\zeta\left(3\right)$ is the Riemann zeta function with argument 3, also known as
the Ap\'ery's constant and $\gamma$ is the Euler - Mascheroni constant. On the other hand,
these quantities for a Fr\'echet distribution are given by:
\begin{eqnarray}
E[X] &=& \mu + \beta\frac{\left({\cal{G}}_{1}\!\left(\xi\right) - 1\right)}{\xi} \\
\mathrm{Var}(X) &=& \beta^{2}\, \frac{{\cal{G}}_{2}\!\left(\xi\right) - {\cal{G}}_{1}\!\left(\xi\right)^{2}}{\xi^2} \\
\tilde{\mu}_3 &=& \frac{{\cal{G}}_{3}\!\left(\xi\right)
    - 3\,{\cal{G}}_{1}\!\left(\xi\right){\cal{G}}_{2}\!\left(\xi\right)
    + 2\,{\cal{G}}_{1}\!\left(\xi\right)^3}
       {\left({\cal{G}}_{2}\!\left(\xi\right) - {\cal{G}}_{1}\!\left(\xi\right)^2\right)^{3/2}} \\
{\cal{K}}[X] &=&  \frac{{\cal{G}}_{4}\!\left(\xi\right)
         - 4\,{\cal{G}}_{3}\!\left(\xi\right){\cal{G}}_{1}\!\left(\xi\right)
         - 3\, {\cal{G}}_{2}\!\left(\xi\right)^2}
         {\left({\cal{G}}_{2}\!\left(\xi\right) - {\cal{G}}_{1}\!\left(\xi\right)^2\right)^{2}} \nonumber \\
         &+& \frac{\left(\xi\right)^{2}{\cal{G}}_{2}\!\left(\xi\right)
         - 6\, {\cal{G}}_{1}\!\left(\xi\right)^{4}}
         {\left({\cal{G}}_{2}\!\left(\xi\right) - {\cal{G}}_{1}\!\left(\xi\right)^2\right)^{2}} \\
H(X) &=& \log\left(\beta\right) + \gamma\left(\xi + 1\right) + 1
\end{eqnarray}
here, ${\cal{G}}_{k}\!\left(\xi\right) = \Gamma\left(1 - k\xi\right)$, with $k \in \mathbb{N}$ and
$\Gamma$ is the gamma function.

These quantities are computed by using the following expressions for the binned data:
\begin{eqnarray}
E[X] &=& \sum_{n=1}^{N} x_{n}P\left(x_{n}\right) \\
\mathrm{Var}(X) &=& E\left[\left(X - E\left(X\right)\right)^{2}\right] \\
\tilde{\mu}_{3} &=& \frac{E\left[\left(X - E\left(X\right)\right)^{3}\right]}
{\left\{E\left[\left(X - E\left(X\right)\right)^{2}\right]\right\}^{3/2}} \\
{\cal{K}}[X] &=& \frac{E\left[\left(X - E\left(X\right)\right)^{4}\right]}
{\left\{E\left[\left(X - E\left(X\right)\right)^{2}\right]\right\}^{2}} - 3\\
H(X) &=& E\left[-\log\left(P\left(X\right)\right)\right]
\end{eqnarray}


\begin{thebibliography}{99}
\bibitem{comment} GS energies are of course the minima of the spectral sequence, these are the most extreme values. The point we are trying to make in the present work may also be taken as an argument which allows us to treat GS energies as extreme values, in the sense of statistics.
\end{thebibliography}
\end{document}